\documentclass[usenatbib]{mnras}
\usepackage{newtxtext,newtxmath}
\usepackage[T1]{fontenc}
\usepackage{graphicx} 

\usepackage{hyperref}
\usepackage{booktabs}
\usepackage{xcolor}
\hypersetup{
    colorlinks=true,    
    urlcolor= blue,
    citecolor = blue,
    linkcolor = blue
    }

\newcommand{\reviewresponse}[1]{{\color{black}#1}}

\begin{document}
\title{Cometary ion dynamics at a weakly outgassing comet}
\author[V. Steinwand et al.]{
V. Steinwand,$^{1}$\thanks{E-mail: victor.steinwand20@imperial.ac.uk}
P. Stephenson,$^{2}$
Z. M. Lewis,$^{3}$
E. Kallio,$^{4}$
A. Beth,$^{1}$
and M. Galand $^{1}$
\\
$^{1}$Department of Physics, Imperial College London, London, UK\\
$^{2}$ Lunar and Planetary Laboratory, Tucson, AZ, USA\\
$^{3}$ Physics Department, Lancaster University, Bailrigg, UK\\
$^{4}$ School of Electrical Engineering, Aalto University, Helsinki, Finland\\
}
\date{Accepted XXX. Received YYY; in original form ZZZ}
\maketitle

\begin{abstract}
    The ESA/\textit{Rosetta} mission escorted comet 67P/Churyumov-Gerasimenko for two years, exploring its plasma environment across diverse outgassing conditions. Plasma density observations from the Rosetta Plasma Consortium (RPC) are broadly categorized into two regimes for the ion dynamics, linked to the presence of a diamagnetic cavity at \textit{Rosetta}'s location. With a diamagnetic cavity present, ions detected by \textit{Rosetta} are accelerated with respect to the neutral coma. Without a diamagnetic cavity present, at lower outgassing, and nearer the nucleus, ions co-move with the neutrals. We examine the transition between regimes following \textit{Rosetta}'s last detection of the cavity in February 2016. During this transition, global 3D plasma models of the cometary ionosphere underestimate plasma densities. To investigate this underestimation, we assess the sensitivity of cometary ion densities to different parameters using a 3D collisional ion test particle model, driven by electromagnetic fields from hybrid modeling. We show that considering cometary electron cooling is necessary to model cometary ion dynamics within 100 km of the surface. Electron temperatures derived from collisional electron modeling affect ion dynamics via the ambipolar electric field, increasing ion number densities. We further show that the cometary electron cooling exobase organizes \textit{Rosetta} plasma density observations; different ion dynamics regimes are linked to the position of \textit{Rosetta} relative to the exobase. These findings demonstrate that \textit{Rosetta} was below this exobase for much of the post-perihelion period. They justify the absence of ion acceleration in plasma density assessments and the use of uniform electron-impact ionization frequencies between \textit{Rosetta} and the surface during post-perihelion.
\end{abstract}


\section{Introduction}
\label{sec:introduction}


\begin{figure*}
    \centering
    \includegraphics[width=0.75\linewidth]{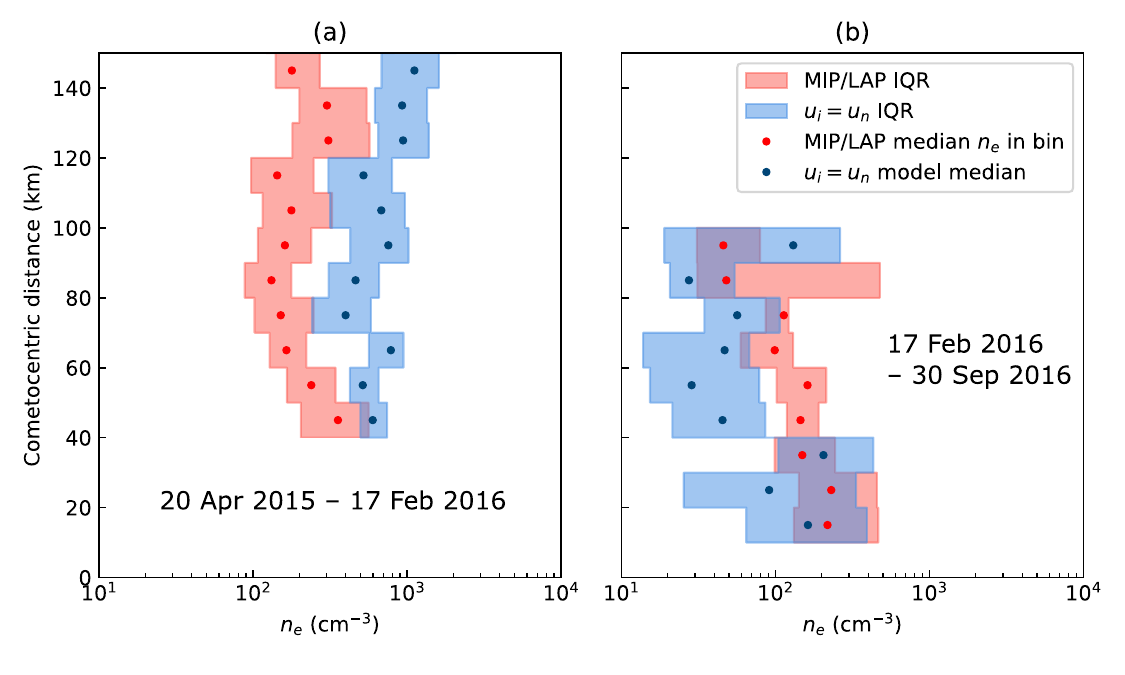}
    \caption{A comparison between modeled cometary ion number densities and measured RPC-MIP/LAP electron number densities for (a) 20 April 2015 to 17 February 2016 and (b) 17 February 2016 to end of mission. Electron number density data from MIP/LAP were binned radially and are shown in red. Bins in blue contain values of $n_i$ calculated using Equation \eqref{eqn:continuity_soln} with $u_i=u_n$ for each neutral number density data point $n_n$ from the ROSINA-COPS instrument. Error bars show the interquartile range (IQR) of the data in each bin. There is better agreement in the time period after perihelion (b) than in the time period where the diamagnetic cavity was present (a).}
    \label{fig:rosetta_data}
\end{figure*}

\begin{figure*}
    \centering
    \includegraphics[width=0.9\linewidth]{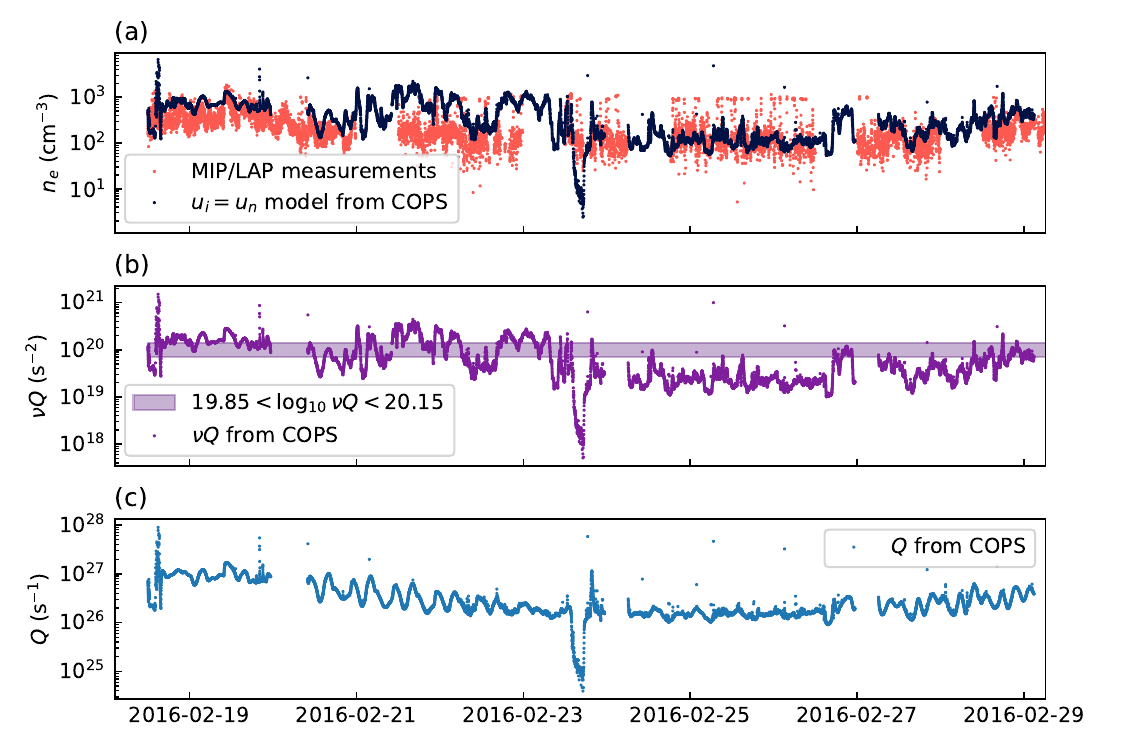}
    \caption{(a) Modeled and measured electron number density $n_e$; (b) total ion production rate $\nu Q$, with the band in $\nu Q$ we consider to be the transition between regimes indicated; (c) and outgassing $Q$ across time. The band of $\nu Q$ chosen corresponds to conditions immediately following the last detection of the diamagnetic cavity. The $u_{i}=u_{n}$ model overestimates number densities during this transition. Maneuvers and dust impacts on the ROSINA-COPS instrument are visible in the time series but had a negligible impact on statistics in this paper. The cometocentric distance of \textit{Rosetta} during this time period is shown in at the start of the time series in Figure \ref{fig:exobase_loc}.}
    \label{fig:time_series}
\end{figure*}


The cometary ionosphere forms by the partial ionization of the neutral gas envelope, or coma, surrounding the cometary nucleus. Neither the coma nor the ionosphere is bound by the weak gravity of the comet. Both expand into space, with the neutral gas continuously replaced by fresh sublimation of ice from the surface of the nucleus. Extreme ultraviolet (EUV) solar photons and energetic (>12--15 eV) electrons then ionize this gas, replenishing the plasma of the ionosphere \citep{galand2016}. The \textit{Rosetta} mission escorted comet 67P/Churyumov-Gerasimenko for 2 years \citep{glassmeier2007a, taylor2017}, providing insight into the cometary plasma environment over a range of different conditions with the instruments of the Rosetta Plasma Consortium \citep[RPC;][]{carr2007}. Plasma number densities were measured by the Mutual Impedance Probe \citep[RPC-MIP;][]{trotignon2007} and the Langmuir Probe \citep[RPC-LAP;][]{eriksson2007}, while electron energy spectra were measured by the Ion and Electron Sensor \citep[RPC-IES;][]{burch2007}. The magnetic field was measured by the fluxgate magnetometer \citep[RPC-MAG;][]{glassmeier2007}. The neutral atmosphere was probed by the Rosetta Orbiter Spectrometer for Ion and Neutral Analysis \citep[ROSINA;][]{balsiger2007}, with the neutral number density measured by the Cometary Pressure Sensor (ROSINA-COPS), corrected for ion composition by the Double Focusing Mass Spectrometer (ROSINA-DFMS).

The long escort mission enabled characterization of the plasma environment of comet 67P across different physical regimes. Previous studies of cometary ions using \textit{Rosetta} observations categorized the ion dynamics into two separate regimes, distinguished by the presence of the diamagnetic cavity at the location of \textit{Rosetta}. The diamagnetic cavity is a region of space around the comet free of magnetic field, which was detected when 67P was around perihelion \citep{goetz2016b, goetz2016a}. With a diamagnetic cavity present, the ions were affected by the comet-solar wind interaction when \textit{Rosetta} was above the boundary \citep{rubin2015, beth2017a}, and were observed to have been radially accelerated \citep{vigren2017b} by the ambipolar electric field \citep{lewis2024} when \textit{Rosetta} was below the boundary. Without the diamagnetic cavity present, when the spacecraft was closer to the comet, the ions detected by \textit{Rosetta} were co-moving with neutral gas \citep{galand2016, vigren2019}.

The physics of the cometary ion population can be investigated using the cometary ion number density. For a weakly outgassing comet, where dissociative recombination and photoabsorption are negligible \citep{galand2016, heritier2018, beth2019}, the cometary ion number density is the solution to the continuity equation:
\begin{equation}
    \frac{\partial n_i}{\partial t} + \nabla\cdot(n_i \mathbf{u}_i) = \nu n_n
    \label{eqn:continuity}
\end{equation}
where $n_i$ is the ion number density, $\mathbf{u}_i$ is the ion bulk velocity, $\nu$ is the total ionization frequency (including both photoionization and electron-impact ionization), and $n_n$ is the neutral number density \citep{beth2022}. At \textit{Rosetta} for low outgassing the attenuation of the ionizing electron flux is negligible; ionization frequencies are spatially uniform \citep{heritier2017a}. In steady-state, for radially moving ions \citep[justified in][]{galand2016}, the solution is therefore given analytically \citep{galand2016, heritier2017a, heritier2018} by
\begin{equation}
    n_i(r) = \frac{\nu n_n(r)(r-r_c)}{u_i(r)}=\frac{\nu Q (r-r_c)}{4 \pi u_i(r) u_n r^2}
    \label{eqn:continuity_soln}
\end{equation}
 where $u_n$ is the radial speed of the neutral gas, $u_i$ is the radial speed of the cometary ions, $r$ is the cometocentric distance, and $r_c$ is the radius of the nucleus. The neutral density $n_n=\nu Q/4\pi u_n r^2$ is described using a Haser model \citep{haser2020}, where $Q$ is the rate at which neutral gas particles sublimate off the nucleus. \reviewresponse{While \cite{haser2020} includes depletion by ionization, it is negligible at the distances probed by \text{Rosetta}; observations are consistent with $n_n \propto 1/r^{2}$ \citep{bieler2015, hassig2015}}.

A comparison between the plasma number densities from consolidated RPC-MIP/LAP measurements \citep{johansson2021} and those calculated using Equation \eqref{eqn:continuity_soln} with $u_i=u_n$ (from here referred to as the $u_i=u_n$ model) is shown for two time periods in Figure \ref{fig:rosetta_data}. Figure \ref{fig:rosetta_data}(a) shows the comparison for the time period around perihelion, from the first detection of the diamagnetic cavity 20 April 2015 to the last detection on 17 February 2016. Figure \ref{fig:rosetta_data}(b) shows the comparison for the post-perihelion period, from 17 February 2016 to end-of-mission. The $u_i=u_n$ model is driven by data using a multi-instrument approach \citep{galand2016, heritier2018}. Local neutral number densities were from ROSINA-COPS, corrected for neutral composition \citep{gasc2017}. Photoionization frequencies were taken from \cite{stephenson2023}, calculated using solar fluxes observed at 1AU by TIMED-SEE \citep{woods2005} and extrapolated to comet 67P. Electron impact ionization frequencies were calculated from RPC-IES electron differential fluxes \citep{burch2007}, corrected for the spacecraft potential as described in \cite{heritier2018}; the electron-impact ionization frequencies we use over the mission were derived by \cite{stephenson2023}.

During the time period around perihelion (at high outgassing with a diamagnetic cavity), Figure \ref{fig:rosetta_data}(a) shows that the $u_i=u_n$ model overestimates ion number densities. From Equation \eqref{eqn:continuity_soln}, an overestimate indicates acceleration of the ions. When a diamagnetic cavity is present, the ions decouple from the neutrals and undergo acceleration before reaching \textit{Rosetta} \citep{lewis2024}. From Figure \ref{fig:rosetta_data}(b), in the post-perihelion period (at low outgassing and in the absence of a diamagnetic cavity) ions move radially at the speed $u_n$ of the neutral gas \citep{galand2016, heritier2018}. \cite{vigren2019} and \cite{lewis2025} showed the $u_i=u_n$ model to be an accurate description for the time period from March 2016 to the end of the mission. The organization of the data over cometocentric distance is revisited in Section \ref{sec:comparison_to_data} (Fig. \ref{fig:electron_exobase_demo}).

We seek to understand the transition between these two regimes in the roughly two weeks from 17 February 2016 to 1 March 2016 by studying the state of the comet when the total ion production rate, defined as the ionization frequency $\nu$ times the local neutral gas production rate $Q$, sits in the band $19.85\leq \log_{10}(\nu Q [\text{s}^{-2}]) \leq 20.15$. As shown in Figure \ref{fig:time_series}, this describes the comet in the days immediately after the last detection of the diamagnetic cavity, while the $u_i = u_n$ model overestimates ion number densities (see Fig. \ref{fig:time_series}a). Plasma models of this transition should be able to reconcile ion acceleration via the ambipolar electric field with the observed lack of acceleration at low outgassing.

Models including acceleration must be able to explain observations under low outgassing conditions, which are consistent with the $u_i=u_n$ model at the location of \textit{Rosetta}. They must also include magnetic fields, as the diamagnetic cavity is no longer present. The low neutral densities, large mean free paths, and non-radial motion introduced by the interaction of the cometary plasma with the solar wind make ion fluid approaches unsuitable as global models of weakly outgassing comets. A kinetic treatment is required to describe the solar wind-cometary plasma interaction. \reviewresponse{Approaches with kinetic ions include particle-in-cell (PiC) models, which also treat electrons kinetically \citep[e.g.][]{deca2017, divin2020}. They are collisionless and therefore usually applied at lower outgassing. There are also hybrid models, which treat electrons as a fluid. Hybrid models have been shown to underestimate ion number densities during this transition period \citep{moeslinger2024, lewis2025}}; the ions are more strongly accelerated in hybrid models than in reality. \reviewresponse{As ion-scale dynamics are consistent between hybrid and PiC models} \citep{deca2019}, this shows that the physical assumptions underlying plasma models require examination. 

In this study, we consider two physical processes which can reduce the acceleration of ions: (1) collisions of ions with the neutral background, and (2) more realistic estimation of the ambipolar electric field, which has been shown to be the main mechanism of ion acceleration close to the cometary nucleus \citep{madanian2016, lewis2025}. The ambipolar electric field is a component of the generalized Ohm's law and has the form \citep{cravens2004}
\begin{equation}
    \mathbf{E}_{ambi} = -\frac{\nabla \mathcal{P}_e}{en_e} = - \frac{k_{B}\nabla (n_e T_e)}{en_e}.
    \label{eqn:ambi}
\end{equation}
where $\mathcal{P}_{e}$ is the electron pressure, $n_{e}$ is the electron number density, $T_{e}$ is the electron temperature, $e$ is the electron charge, and $k_{B}$ is the Boltzmann constant. It has been shown to trap electrons in regions of denser neutral gas close to the nucleus, increasing their collisionality and creating a population of cold electrons \citep{stephenson2022, stephenson2024}, which were observed at low outgassing at 67P \citep[e.g.][]{eriksson2017, engelhardt2018, wattieaux2019, gilet2020}. Electron trapping is the main explanation for the cold electron population observed by \textit{Rosetta} when the coma was too thin to be collisional for radially moving electrons \citep{engelhardt2018}. This effect on electron temperature feeds back onto the ambipolar electric field according to Equation \eqref{eqn:ambi}, flattening its potential well \citep{stephenson2024}. In contrast, three-dimensional plasma models often assume adiabatic electrons \citep[e.g.][]{koenders2013, alho2019, gunell2024} and neglect the increase in cometary electron collisionality along with the resulting feedback mechanism on the ambipolar potential.

We have examined the sensitivity of a three-dimensional ion test particle model, using electromagnetic fields from a hybrid model, to both ion-neutral collisions (including ion-neutral chemistry) and realistic modeling of the ambipolar electric field. In Section \ref{sec:methods}, we explain the method used in our simulation and the present key parameters and assumptions. In Section \ref{sec:sim_results}, we examine the results of the simulation on their own to understand the physics involved and the sensitivity of the ion number densities to ion-neutral collisions and the electron temperature via the ambipolar electric field. This is followed in Section \ref{sec:comparison_to_data} by a comparison of the simulations to \textit{Rosetta} plasma data, and a qualitative discussion of the physical consequences of collisional electrons as observed in the data over the post-perihelion period.

\section{Methods}
\label{sec:methods}

\subsection{Models in this study}

\begin{table}
    \centering
    \begin{tabular}{c|c|c|c}
    \toprule
      \multicolumn{4}{c}{\textbf{Ion test particle model cases}}\\
      \multicolumn{2}{c}{Case} & Collisional? & $T_e$ \\\midrule
      \multicolumn{2}{c}{1} & \multicolumn{1}{c}{No}
        & \multicolumn{1}{c}{Adiabatic} \\
      \multicolumn{2}{c}{2} & \multicolumn{1}{c}{Yes}
        & \multicolumn{1}{c}{Adiabatic} \\ 
        
      \multicolumn{2}{c}{3} & \multicolumn{1}{c}{No}
        & \multicolumn{1}{c}{Imposed} \\
      \multicolumn{2}{c}{4} & \multicolumn{1}{c}{Yes}
        & \multicolumn{1}{c}{Imposed} \\\midrule
        
      \multicolumn{4}{c}{\textbf{Ion test particle model parameters}}\\
      \multicolumn{2}{c}{Parameter} & \multicolumn{2}{c}{Value}\\\midrule
      \multicolumn{2}{l}{Total ion production$^\dagger$ $(\nu Q)$} & \multicolumn{2}{c}{$10^{20}\text{ s}^{-2}$} \\
      \multicolumn{2}{l}{Neutral gas speed$^\dagger$ $(u_n)$} & \multicolumn{2}{c}{$700 \text{ m s}^{-1}$} \\
      \multicolumn{2}{l}{Total ionization frequency ($\nu$)} & \multicolumn{2}{c}{$2 \times 10^{-7} \text{ s}^{-1}$}\\
      \multicolumn{2}{l}{Outgassing ($Q$)} & \multicolumn{2}{c}{$5 \times 10^{26} \text{ s}^{-1}$}\\
      \multicolumn{2}{l}{$\text{NH}_3$ fraction of neutral gas} & \multicolumn{2}{c}{0.002}\\\\
      \multicolumn{3}{l}{$^\dagger$ parameter shared with hybrid model} \\
      \midrule

      \multicolumn{4}{c}{\textbf{Hybrid model parameters}}\\
      
      Species   &  \multicolumn{3}{c}{\hspace{10pt}Parameter}\\
      & $n$ & $u$ & $T$ \\\midrule
      Solar wind $\text{H}^+$ & $1 \text{ cm}^{-3}$ & $400 \text{ km s}^{-1}$ & $61000 \text{ K}$ \\
      (upstream)\\
      Solar wind $\text{He}^{++}$ & $0.05 \text{ cm}^{-3}$ & $400 \text{ km s}^{-1}$ & $214000 \text{ K}$\\
      (upstream)\\
      Solar wind $e^{-}$ & $1 \text{ cm}^{-3}$ & $400 \text{ km s}^{-1}$ & $139254 \text{ K } (12 \text{ eV})$\\
      (upstream) \\
      Cometary $\text{H}_2\text{O}^+$ & (modeled) & (modeled) & $6000 \text{ K}$ (initial) \\\midrule
      \multicolumn{2}{c}{Parameter} & \multicolumn{2}{c}{Value}\\\midrule
      \multicolumn{2}{c}{Interplanetary magnetic} & \multicolumn{2}{c}{3 nT} \\
      \multicolumn{2}{c}{ field (IMF) magnitude}

    \\\bottomrule
    \end{tabular}
    \caption{Ion test particle model parameters and the four cases used in the sensitivity study. Parameters are constant across all cases. The adiabatic and imposed electron temperatures are given by equations \eqref{eqn:adiabatic} and \eqref{eqn:analytical_fit}, respectively; the temperature profiles across cometocentric distance are shown in Figure \ref{fig:Te} along with the resulting ambipolar electric fields. Upstream quantities driving the hybrid model are imposed as boundary conditions -- the upstream electron temperature was equal for both the imposed electron temperature profile and the adiabatic profile.}
    \label{tab:params}
\end{table}

\begin{figure}
    \centering
    \includegraphics[width=\linewidth]{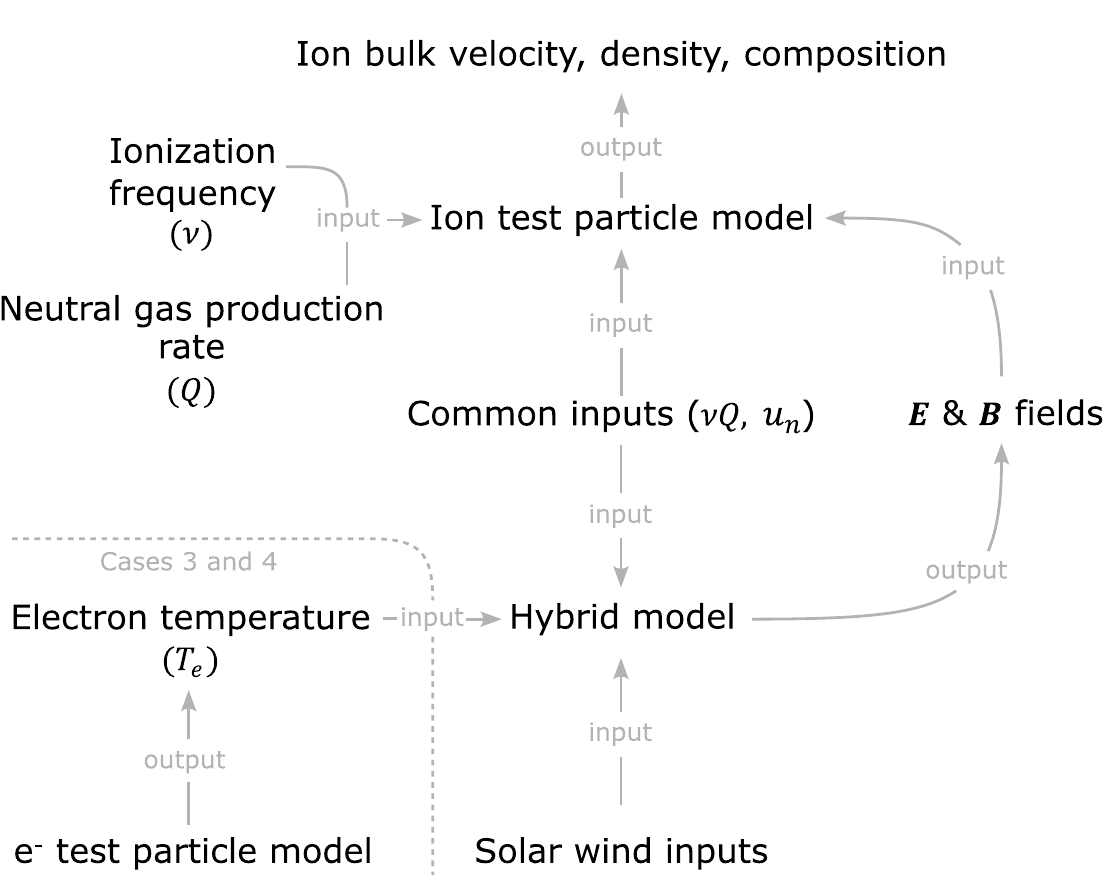}
    \caption{A schematic describing the interaction of the different models and quantities used in this study. The electron test particle model is only used as an input to the hybrid model for the cases where collisional electrons are considered.}
    \label{fig:model_schematic}
\end{figure}

We have used the three-dimensional collisional ion test particle model presented in \cite{lewis2025} to calculate densities and bulk velocities for three ion species ($\text{H}_2\text{O}^+$, $\text{H}_3\text{O}^+$, $\text{NH}_4^+$) in a cometary ionosphere. Electromagnetic fields were provided by a hybrid model \citep{simonwedlund2017, alho2019}, where the ions are simulated as individual particles and the electrons are treated as a massless fluid. The relationships between models and their input and output quantities are shown in Figure \ref{fig:model_schematic}. We \reviewresponse{use} different runs of the ion test particle model to assess the sensitivity of the simulated densities to two parameters: ion collisionality, slowing the ions directly, and electron collisionality, affecting the ions via the electron temperature and the ambipolar electric field (Eq. \ref{eqn:ambi}). We examine a total of four cases, for two different levels of ion collisionality and two different electron temperature profiles. Numerical parameters for each model and case are shown in Table \ref{tab:params}.

The effect of electron collisionality was assessed via its impact on the ambipolar electric field. While the ambipolar field has been shown to be the strongest driver of acceleration close to the comet \citep{madanian2016, lewis2025}, in a physical regime with no diamagnetic cavity, magnetic fields cannot be neglected, motivating the use of a hybrid model rather than an analytical expression as in \cite{lewis2024}. We produced two sets of fields with the hybrid model: one where the electron temperature is treated adiabatically and one with an electron temperature imposed, as discussed in Section \ref{sec:methods:Te}. Cases 1 and 2 have adiabatic electrons (no electron collisions) and cases 3 and 4 have an imposed electron temperature (collisional electrons).  \reviewresponse{The ability to easily impose an electron temperature requires the use of a hybrid model over a PiC model}.

The ion test particle model includes collisions between the ions and the neutral background. While the hybrid model includes Langevin drag, which consists of a continuous, non-stochastic frictional term in the ion equation of motion \citep{cravens1987, simonwedlund2017, alho2019}, also used in fluid models \citep{gombosi1996, rubin2014a}, this is insufficient to assess the impact of ion-neutral collisionality; the test particle model has the advantage of including fully stochastic momentum transfer collisions as well as ion-neutral chemistry \citep[cross-section given in Appendix B of][]{lewis2025}.

\begin{figure*}
    \centering
    \includegraphics[width=0.75\linewidth]{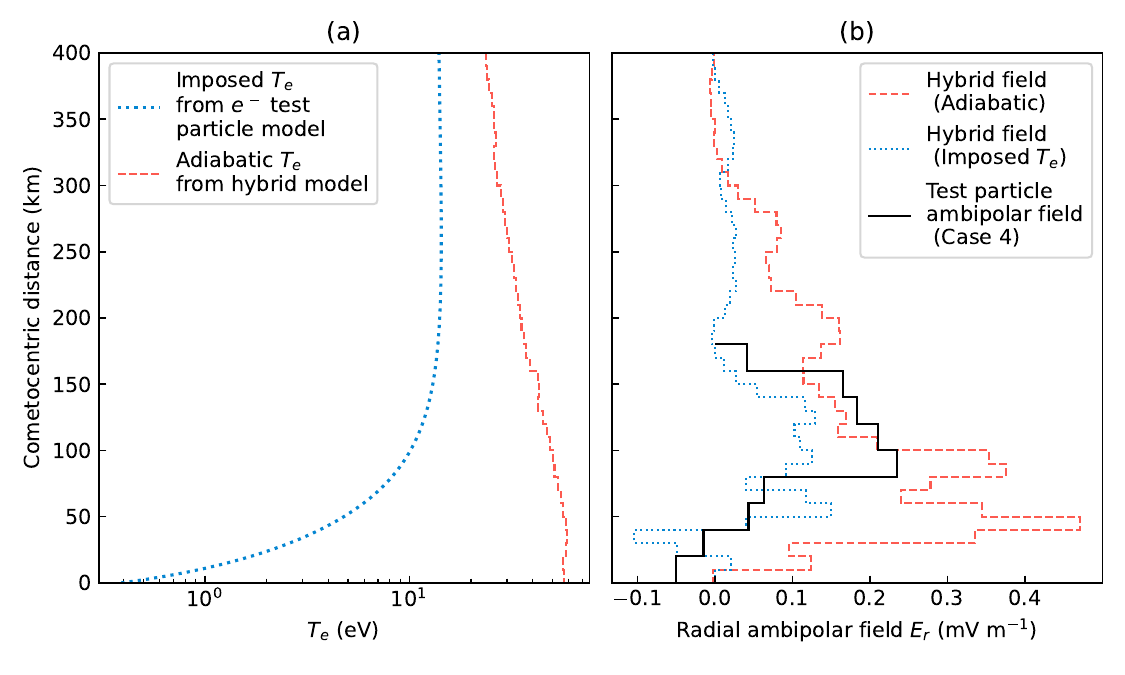}
    \caption{(a) Electron temperatures and (b) ambipolar electric fields, averaged across radial bins in the terminator plane (which was probed by \textit{Rosetta}). The ambipolar field calculated from the total ion number density produced by Case 4 as described in Table \ref{tab:params} is shown, and is qualitatively consistent with the field produced by the hybrid model; Case 4 best described the cometary conditions during the transition between physical regimes immediately after the final detection of the diamagnetic cavity.}
    \label{fig:Te}
\end{figure*}
While both models consider the production of $\text{H}_2\text{O}^+$ via photoionization and electron-impact ionization, the ion test particle model also includes collisions resulting in proton transfer, producing the  cometary ion species $\text{H}_3\text{O}^+$ and $\text{NH}_4^+$ \citep{altwegg1993}, which the hybrid model neglects entirely. Cometary ion species that are produced via ion-neutral chemistry, such as $\text{H}_{3}\text{O}^+$ and $\text{NH}_4^+$ can dominate the composition of the ionosphere \citep{vigren2013, beth2017}; since ion-neutral chemistry decelerates the overall ion population, it is necessary to include its effect in the collisional model. Proton transfer to $\text{NH}_3$ produces $\text{NH}_4^+$, and $\text{NH}_3$ has the highest proton affinity of common cometary neutral species \citep{beth2022}, making it a good probe for the effect of ion-neutral chemistry; for this reason the ion test particle model includes $\text{NH}_4^+$ and the $\text{H}_2\text{O}$-dominated neutral background in the model includes $\text{NH}_3$, while we neglect other constituents of the neutral coma (e.g. $\text{CO}_2$).

To assess the impact of collisions, collisionless runs (cases 1 and 3) were compared to collisional runs (cases 2 and 4) of the ion test particle model for the different electron temperature cases. Each collisionless run had the same inputs as each collisional run with all collision cross sections set to 0 (see Table \ref{tab:params}).

\subsection{Electron temperatures}
\label{sec:methods:Te}
In cases 1 and 2 (see Table \ref{tab:params}), the electron temperature $T_e$ is adiabatic, defined as
\begin{equation}
    T_e^{adia.} = T_{e, SW} \left(\frac{n_e}{n_{e, SW}}\right)^{\gamma-1}
    \label{eqn:adiabatic}
\end{equation}
where $n_e$ is the local electron number density; $n_{e,SW}$ and $T_{e, SW}$ are the upstream solar wind electron number density and temperature, respectively. This is calculated self-consistently with respect to the local electron number density in the hybrid model.

\begin{figure*}
    \centering
    \includegraphics[width=0.99\linewidth]{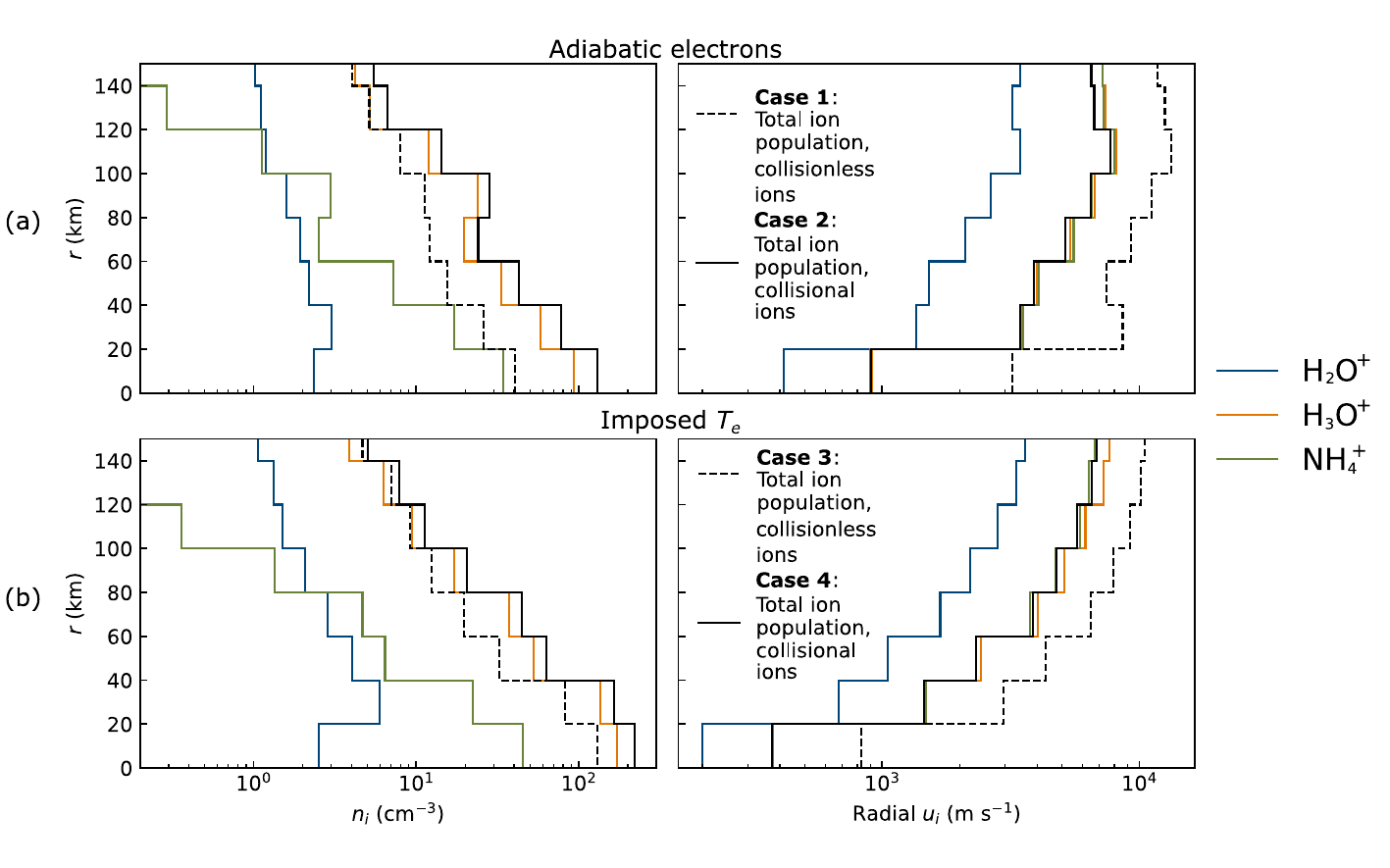}
    \caption{Ion densities (left column) and bulk radial speeds (right column) for each species for (a) adiabatic electrons and (b) the imposed electron temperature profile. The only species shown for the collisionless case is $\text{H}_2\text{O}^+$ (blue) as there was no production of other species via chemistry; however, the collisional model shows that close to the nucleus $\text{H}_3\text{O}^+$ (orange) and $\text{NH}_4^+$ (green) are the dominant ion species. The cometary radius \reviewresponse{was set to 2 km.}}
    \label{fig:composition}
\end{figure*}

In cases 3 and 4, the imposed temperature profile is an analytical fit to the electron temperature found by the collisional electron test particle model of \cite{stephenson2024} run at $\nu=1.12\times10^{-7} \text{ s}^{-1}$ and $Q=5\times10^{26} \text{ s}^{-1}$ and radially averaged in the terminator plane (which was probed by \textit{Rosetta}):
\begin{equation}
    T_e^{imp.} = \frac{p_1 r^2 + p_2 r + p_3}{r^2+q_{1}r+q_2}
    \label{eqn:analytical_fit}
\end{equation}
where the imposed temperature $T_{e}^{imp.}$ has units of electron volts, $r$ is cometocentric distance in kilometers and
\begin{equation}
    \begin{alignedat}{2}
    &p_1 = &11.3896109 \\
    &p_2 = &612.530963 \\
    &p_3 = &5525.67913 \\
    &q_1 = -&63.1739938 \\
    &q_2 = &14203.0527\\
    \label{eqn:analytical_fit_params}
    \end{alignedat}
\end{equation}
are the fitted parameters.

The adiabatic and imposed electron temperature profiles and their respective resulting ambipolar electric fields as calculated by the hybrid model are shown in Figure \ref{fig:Te}. The ambipolar electric field is weakened by imposing an electron temperature close to the comet, but the ambipolar field with $T_e$ imposed converges with the adiabatic field at the simulation boundary at the undisturbed solar wind electron temperature, where the electron population is dominated by solar wind electrons. This is further consistent with the adiabatic behavior of solar wind electrons in particle-in-cell (PiC) models \citep[e.g.][]{deca2019}.

The ambipolar electric field derived from the imposed electron temperature and the total ion number density output of Case 4 of the ion test particle model is shown in black and is discussed in Section \ref{sec:sim_results}. The electron test particle model, like the ion test particle model, requires external input for the electromagnetic fields. In the case of the \cite{stephenson2024} model, the fields were from a fully kinetic, collisionless PiC model \citep{deca2017, deca2019}.

\section{Simulation results and sensitivities}
\label{sec:sim_results}

\begin{figure*}
    \centering
    \includegraphics[width=0.9\linewidth]{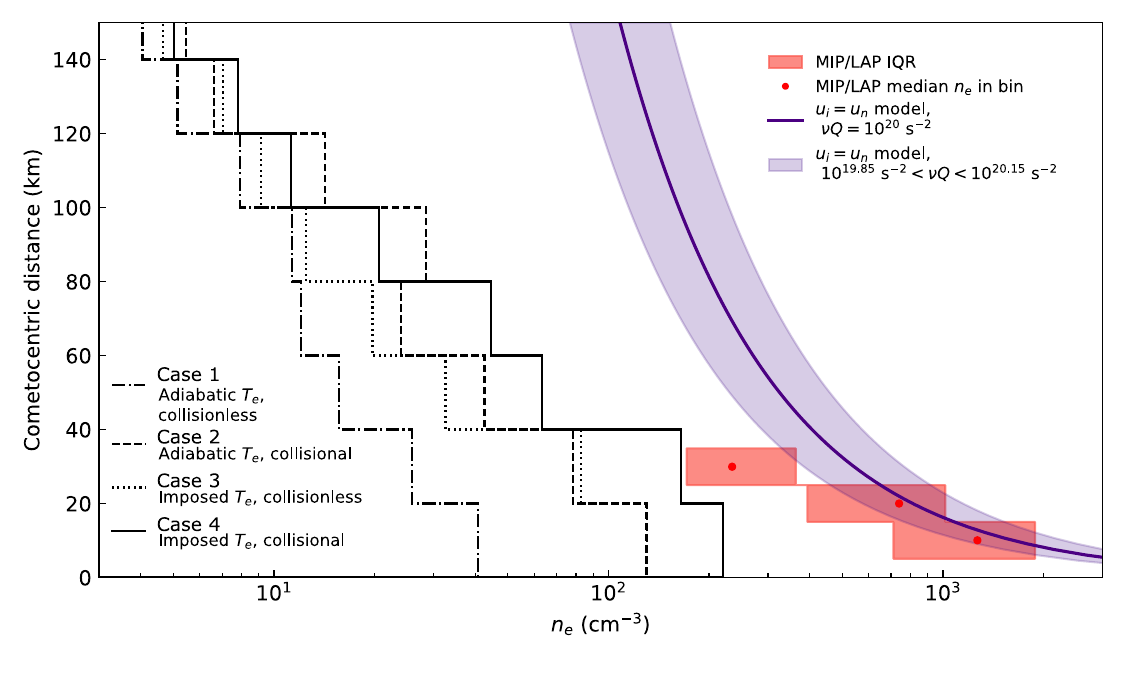}
    \caption{Ion number densities for each of the four cases of the ion test particle model. Model runs are shown in black. RPC-MIP/LAP data is shown in red, binned by cometocentric distance. The error region is the interquartile range of the data in each bin. The highest of the three data bins mainly contains data from the two weeks following the last appearance of the diamagnetic cavity on 17 February 2016, with 2949 out of 4433 data points coming from the two-week period; the parameters of this study were chosen to most closely align with this time period. The ion number density produced by the collisional, imposed $T_e$ case of the ion test particle model was the closest to the Rosetta data. }
    \label{fig:sensitivity}
\end{figure*}

The ion number densities and bulk speeds for each case in \mbox{Table \ref{tab:params}} and each ion species are shown in Figure \ref{fig:composition}. Ion number densities and bulk radial speeds for an adiabatic electron temperature and the resulting ambipolar electric field are shown in Figure \ref{fig:composition}(a), and number densities and speeds for the imposed electron temperature are shown in \mbox{Figure \ref{fig:composition}(b)}. The collisionless results for each ambipolar electric field are shown as a dashed line. Total ion densities for all four cases are compared in Figure \ref{fig:sensitivity}. As expected, the highest total ion number densities are exhibited by case 4, shown as the solid black line in Figures \ref{fig:composition}(b) and \ref{fig:sensitivity}. This case included ion collisions and used an imposed ambipolar electric field weakened by electron collisionality, resulting in reduced acceleration and increased ion number densities. These results are also the closest to \textit{Rosetta} data, which is further discussed in Section \ref{sec:comparison_to_data}.

As the input parameters for case 4 are closest to multi-instrument observations, the ambipolar electric field for this case is shown in Figure \ref{fig:Te}. It is calculated using Equation \eqref{eqn:ambi} with the ion number density from the ion test particle model and the imposed electron temperature given in Equation \eqref{eqn:analytical_fit}. The field shows qualitative consistency with the hybrid model. Although the ion test particle model does not have self-consistent feedback between the ambipolar electric field and the ion number density, the consistency between the hybrid and ion test particle fields indicates that the ion number densities are reliable. The inclusion of self-consistent feedback between the ambipolar electric field and the ion number density would not greatly affect the results of the simulation.

\subsection{Sensitivity to ion collisions}
One effect of ion collisionality in Figure \ref{fig:composition} is the ion composition, which the collisionless models are unable to reproduce. The hybrid model and the collisionless runs of the ion test particle model only consider $\text{H}_2\text{O}^+$, while at low cometocentric distances Figure \ref{fig:composition} shows that $\text{H}_3\text{O}^+$ and $\text{NH}_4^+$ are the dominant ion species. The different kinds of collisions also increase the total ion density. Ions are directly decelerated by momentum transfer collisions. The slight differences in mass between $\text{H}_2\text{O}^+$ and $\text{H}_3\text{O}^+$ imply a weaker acceleration of $\text{H}_3\text{O}^+$ ions when affected by the same electric field, resulting in higher total densities. Proton transfer neutralizes an ion that may have previously been accelerated by the ambipolar electric field and replaces it with an ion traveling at the neutral speed $u_n$, also decreasing the cometary ion bulk velocity. The effect of ion collisions is not limited to regions close to the cometary nucleus -- while the number densities converge at large distances for each case of electron collisionality, the bulk speeds consistently remain higher for the cases without ion collisions than for the cases with ion collisions.

\subsection{Sensitivity to electron collisions}

The effect of the electron temperature is most pronounced close to the comet. As shown in Figures \ref{fig:Te} and \ref{fig:composition}, ion number densities are increased and bulk speeds are decreased by imposing an electron temperature rather than treating the electrons adiabatically. The effect is present in both the collisional and collisionless cases. The number density and bulk radial speed profiles both converge beyond a cometocentric distance of 100 km, showing that the effect of the electron trapping and cooling on the ion dynamics is confined to the vicinity of the nucleus. However, the effect extends beyond the region where the electrons are collisional. From collisional electron test particle modeling, the height of the inelastic collision exobase for cometary electrons trapped by the ambipolar electric field can be estimated as \citep{stephenson2024}
\begin{equation}
    \begin{split}
    r_{exo} &= 100 n_n \sigma r^2 \\
    &= \frac{100Q\sigma}{4 \pi u_n}
    \end{split}
    \label{eqn:r_exobase}
\end{equation}
where $\sigma$ is the inelastic collision cross-section of a typical cometary electron at 10 eV, with a value of $6.67 \times 10^{-17} \text{ cm}^{2}$. The dimensionless factor of 100 approximates the longer, three-dimensional path taken by trapped cometary electrons through collisional regions of the coma when compared to purely radial motion. For the parameters used in this simulation, the location of the electron cooling exobase was at $\sim38\text{ km}$, showing that collisional electrons affect ion dynamics beyond the cometary electron cooling exobase.

\section{Comparison to \textit{Rosetta} data}
\label{sec:comparison_to_data}

\begin{figure*}
    \centering
    \includegraphics[width=0.58\linewidth]{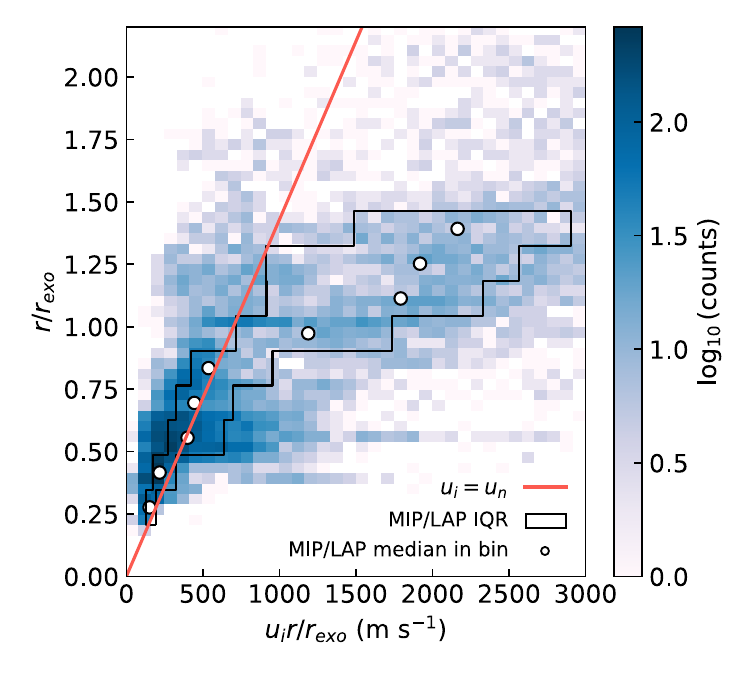}
    \caption{Two-dimensional histograms of post-perihelion MIP/LAP datapoints from 17 February 2016 to 30 September 2016, for $19.85\leq \log_{10}(\nu Q [\text{s}^{-2}]) \leq 20.15$, showing the organization of the data in the parameter space $r/r_{exo}$ against $u_ir/r_{exo}=(\nu/100\sigma n_i)(1-r_c/r)$ (Eqn. \ref{eqn:exobase_scaling}). The red boxes and points are values of $(\nu/100\sigma n_i)(1-r_c/r)$ binned in $r/r_{exo}$. The straight line represents the $u_i=u_n$ model. As shown by Equation \eqref{eqn:exobase_scaling}, the local gradient in this parameter space gives the local velocity $u_{i}$ at $r/r_{exo}$; below the exobase, the data are clustered around the $u_i=u_n$ line, showing the utility of the acceleration-free model inside the flattened region of the ambipolar potential well. They depart from the line above $r\sim r_{exo}$, indicating acceleration.}
    \label{fig:electron_exobase_demo}
\end{figure*}

\begin{figure*}
    \centering
    \includegraphics[width=0.9\linewidth]{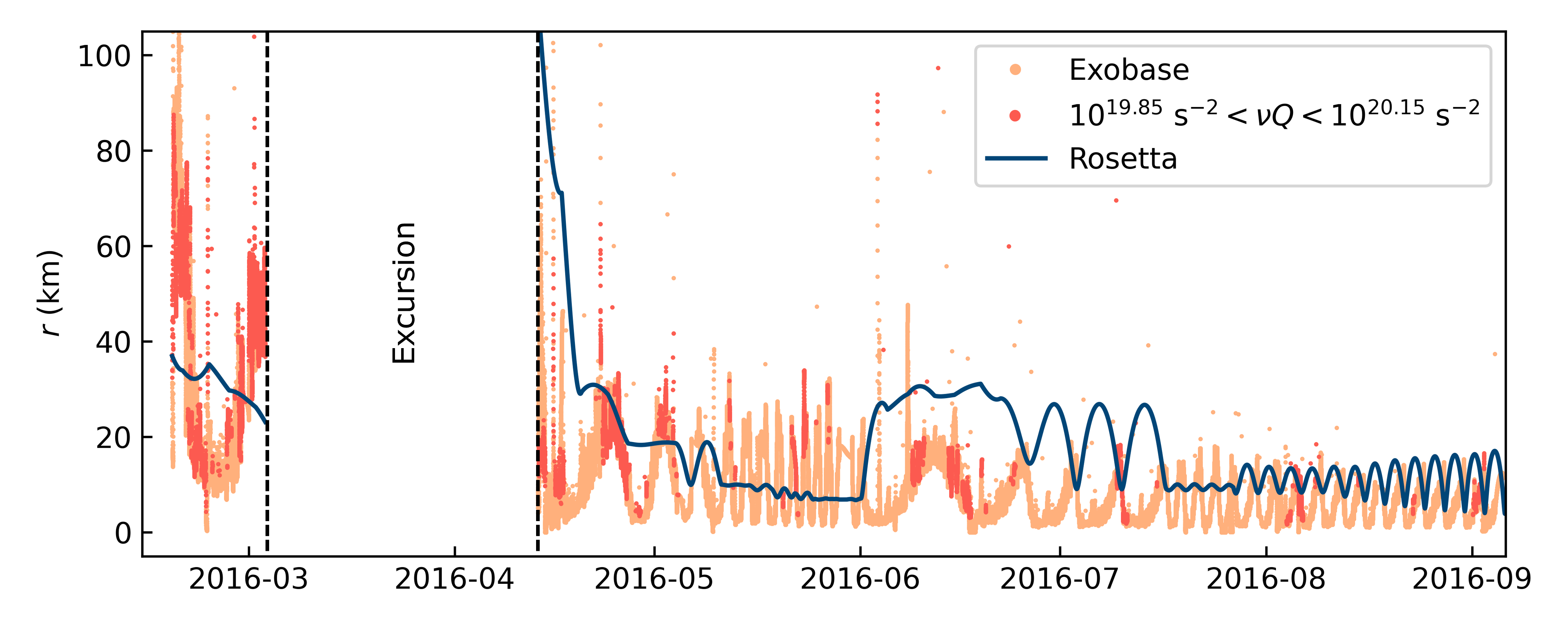}
    \caption{The location of \textit{Rosetta} and the location of the electron cooling exobase calculated from ROSINA-COPS data during the post-perihelion period. Markers in red are exobase locations for the band in $\nu Q$ considered in this study, with the majority of the datapoints above the exobase associated with acceleration coming from the two week transition period. \textit{Rosetta} did not observe cometary ion acceleration at lower levels of outgassing because the spacecraft was near to or below the location of the exobase during this phase of the mission.}
    \label{fig:exobase_loc}
\end{figure*}

As shown in \mbox{Figure \ref{fig:sensitivity}}, an electron temperature derived from collisional electrons brought modeled ion number densities closer to observed number densities when compared to the cases with adiabatic electrons. Case 1, which had neither ion nor electron collisionality, had the lowest ion number density, while case 4, which included both processes, had the greatest number density and produced an ion number density most similar to the \textit{Rosetta} observations. Cases 2 and 3 each included one of the processes, and produced very similar number densities to each other. This shows that ion and electron collisions have similarly important effects on the ion dynamics. Neither is sufficient on its own to explain observed ion number densities.

The ions are moving radially in the bins close to the comet, justifying the use of equations \eqref{eqn:continuity} and \eqref{eqn:continuity_soln} in analyzing number densities and a comparison to the $u_i=u_n$ model. The data are well-explained by the $u_i=u_n$ model up to a cometocentric distance of 25 kilometers for $19.85\leq \log_{10}(\nu Q [\text{s}^{-2}]) \leq 20.15$, departing from the model in the bin above this threshold, which corresponds to the transition time period. The interquartile range (IQR) of the data in the \mbox{25--30 km} bin spans between the $u_i=u_n$ model and case 4 of the ion test particle model. None of the modeled densities for any of the ion test particle cases agrees with observations for cometocentric distances below 25 km; however, our results in this first bin are fundamentally limited by the 7.8 km resolution of the Cartesian grid used by the hybrid model, which sets the resolution of the input electric and magnetic fields for the ion test particle model. \textit{Rosetta} data are already well-explained by the $u_i=u_n$ model in this range of cometocentric distance \citep{heritier2017a} -- the most relevant physical comparison is between the RPC-MIP/LAP data and the \mbox{25--35 km} bin of the ion test particle model.

Despite its quantitative limitations, the ion test particle model reproduces the underlying physical effects qualitatively. The departure from the $u_i=u_n$ model and the trend towards the ion test particle model in the 25--35 km bin can be explained by reorganization of the data with respect to the electron cooling exobase, since the ambipolar potential well is flattened in the region where the electrons experience cooling via inelastic collisions. We can use the estimate of the height of the exobase given by \mbox{Equation \eqref{eqn:r_exobase}} to scale the solution for $n_i$ in Equation \eqref{eqn:continuity_soln}:
\begin{equation}
    n_i = \frac{\nu}{100 \sigma u_i \cdot r/r_{exo}} \left(1-\frac{r_c}{r}\right).
\end{equation}
Rearranging this yields
\begin{equation}
    u_i \frac{r}{r_{exo}} = \frac{\nu}{100 \sigma n_i} \left(1-\frac{r_c}{r}\right).
    \label{eqn:exobase_scaling}
\end{equation}
While Equation \eqref{eqn:r_exobase} makes assumptions such as spherical symmetry and is only a rough estimate, plotting the scaled distance $r/r_{exo}$ against $(\nu/100\sigma n_i)(1-r_c/r)$ does organize \textit{Rosetta} data;  \mbox{Figure \ref{fig:electron_exobase_demo}} shows a histogram in this parameter space for plasma density data points with $n_i$ taken from RPC-MIP/LAP, $\nu$ taken from solar fluxes and RPC-IES, and $r_{exo}$ calculated using ROSINA-COPS data (see Section \ref{sec:introduction}). The data were taken from the post-perihelion time period (see Figure \ref{fig:rosetta_data}) in the total ion production band $19.85\leq \log_{10}(\nu Q [\text{s}^{-2}]) \leq 20.15$. The radius of the nucleus was taken to be 2 km, the typical value for 67P \citep{heritier2017a}. From Equation \eqref{eqn:exobase_scaling}, the local gradient $1/u_i$ reveals the local ion bulk radial speed. The data cluster around a line with a gradient $1/u_i=1/u_n$ for $u_n=700 \text{ m s}^{-1}$ until around $r\sim r_{exo}$, above which the local gradient indicates acceleration. This is consistent with a model of an ambipolar potential well that is flattened below the electron exobase by inelastic collisions, and explains the effectiveness of the $u_i=u_n$ model when \textit{Rosetta} was close to comet 67P at low outgassing.

The cometocentric distance of \textit{Rosetta} and an estimate of the electron cooling exobase is shown over the post-perihelion time period in Figure \ref{fig:exobase_loc}. Red points correspond to the band in ion production $19.85\leq \log_{10}(\nu Q [\text{s}^{-2}]) \leq 20.15$ and are included in the analysis in Figure \ref{fig:electron_exobase_demo}. While the overestimate was found in the transition period, the physical reasoning applies across the whole post-perihelion period. The spacecraft was either near to or below the exobase for the duration of this period. \textit{Rosetta} was therefore too close to the nucleus for any significant acceleration of the cometary ions to have occurred before reaching the spacecraft, which is consistent with observation across low outgassing periods \citep{galand2016, heritier2017a, heritier2018}.

\section{Conclusions}

\begin{figure}
    \centering
    \includegraphics[width=\linewidth]{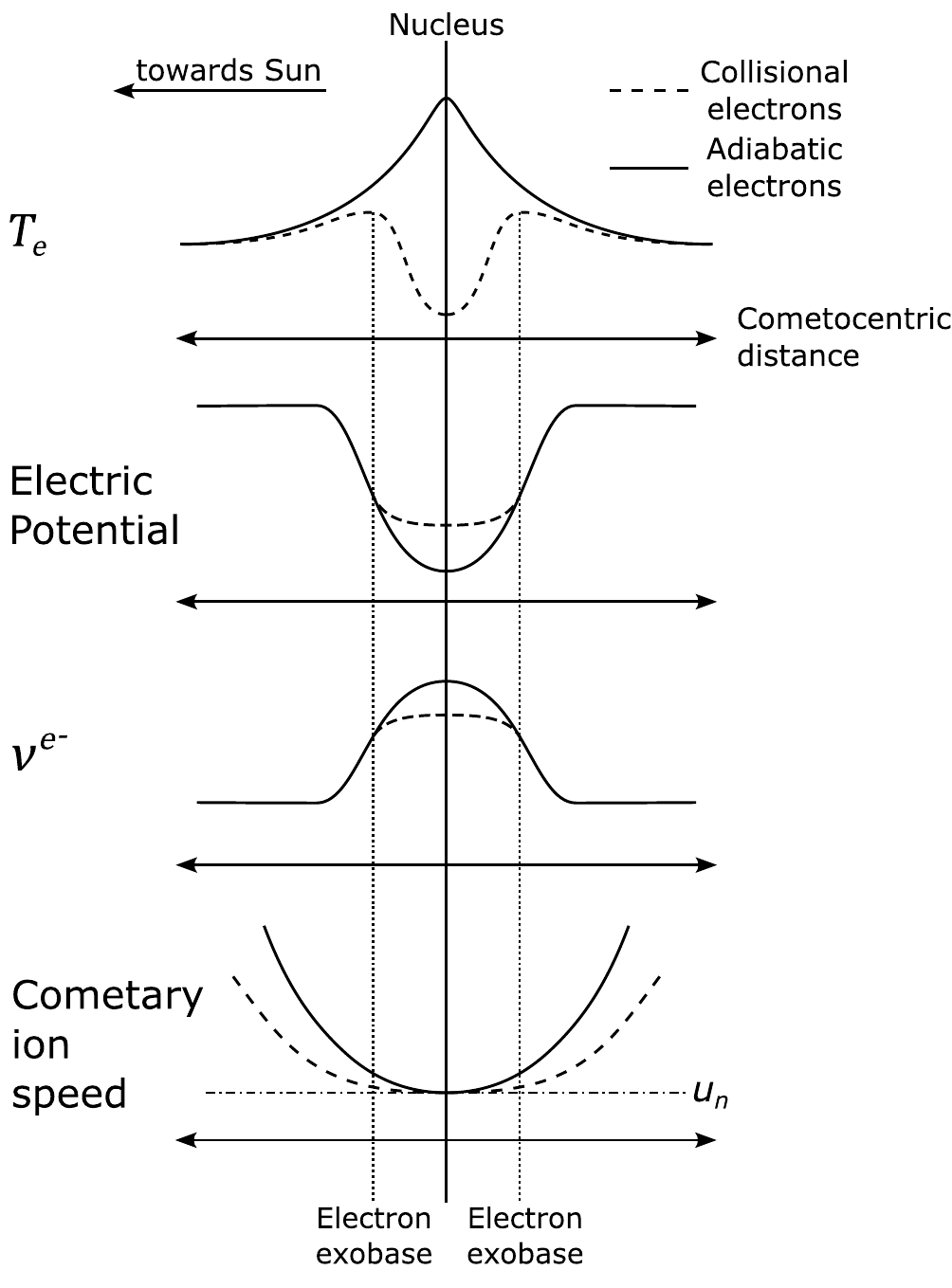}
    \caption{A schematic of various physical quantities at a low outgassing comet as a function of cometocentric distance, comparing the case of collisional and collisionless electrons. The trapping of electrons by the ambipolar electric field leads to enhanced electron collisionality and cooling. The temperature $T_e$ of the overall electron population is reduced by inelastic collisions close to the cometary nucleus. This flattens the ambipolar potential well below the exobase and decreases the strength of the ambipolar electric field. As a result, the effects of the ambipolar electric field are weakened: the acceleration of solar wind electrons into the potential well (which results in electron impact ionization), the acceleration of cometary ions away from the nucleus, and the trapping of the electrons are all reduced. \reviewresponse{The schematic is simplified to illustrate the effect of the ambipolar field and does not capture the inherent asymmetric 3D nature of the comet-solar wind interaction.}}
    \label{fig:exobase_schematic}
\end{figure}

The ambipolar electric field and the feedback loop between the field and trapped electron cooling affect several aspects of cometary physics in the low outgassing regime (\mbox{$\nu Q \approx 10^{20} \text{ s}^{-2}$} and \mbox{$Q\approx 5\times10^{26}\text{ s}^{-1}$} in this study). \reviewresponse{Figure \ref{fig:exobase_schematic} shows a schematic demonstrating the effects of the feedback loop on several processes; while the true interaction of the comet with the solar wind is inherently three-dimensional and asymmetric, the schematic is intentionally simplified to demonstrate the specific ambipolar field effects. Electron cooling feeds back on electron impact ionization frequencies and cometary ion dynamics, two processes that are independently empirically constrained and both consistent with a flat ambipolar electric potential well around the cometary nucleus. }

The ambipolar electric field enables electron impact ionization. The field accelerates solar wind electrons towards the nucleus -- as solar wind electrons fall into the potential well, their kinetic energy increases \citep{deca2017, divin2020, galand2020, stephenson2022}. For a typical solar wind electron at 12 eV, the increase in energy also increases electron impact ionization frequencies \citep{itikawa2005}. \cite{stephenson2023} showed that taking into account the trapping of cometary electrons born in the ambipolar potential well reduces electron impact ionization frequencies. While solar wind electrons are accelerated towards the comet above the electron cooling exobase, the flat portion of the ambipolar potential below the exobase does not further accelerate the solar wind electrons, meaning there is no increase in electron-impact ionization frequencies between the location of \textit{Rosetta} observations and the comet. Furthermore, the neutral densities in this region are too low to significantly attenuate the ionizing electron flux \citep{heritier2017a}, resulting in no significant decrease in electron-impact ionization frequencies. Combining no significant acceleration in the solar wind electrons and no significant energy degradation in the electron flux justifies that there is a constant electron-impact ionization frequency between the exobase and the cometary surface. A constant electron-impact ionization frequency and dissociative excitation frequency below the exobase are consistent with multi-instrument analysis of the electron number density \citep{heritier2017a, heritier2018} and of cometary auroral emissions \citep{galand2020, stephenson2021}.

The ambipolar electric field accelerates ions radially outwards \citep{lewis2025} in the first 100 km above the nucleus; in the sensitivity study in Section \ref{sec:sim_results} we show that the feedback loop between the ambipolar field and the trapped electron cooling weakens the ambipolar electric field and prevents significant acceleration of the ions. In Section \ref{sec:comparison_to_data} we show that this allows the $u_i=u_n$ model \citep[e.g.][]{galand2016, heritier2018, vigren2019} to be effective below the cometary electron cooling exobase, which was probed by \textit{Rosetta} during the post-perihelion time period from February 2016 (see Figure \ref{fig:exobase_loc}). Its effect on both ion and electron dynamics shows that it is critical to account for electron collisional cooling in plasma models for an accurate description of the cometary ionosphere in the first 100 km from the surface of a weakly outgassing comet.

\section*{Acknowledgments}
\reviewresponse{We would like to acknowledge the invaluable work of the RPC team, the \textit{Rosetta} team, and the ESA Planetary Science Archive for providing data.} Work at Imperial College London was supported by the Science and Technology Facilities Council (STFC) of the UK under studentship 2928058 (grant ST/Y509231/1) and grant ST/W001071/1, as well as by the UK Space Agency (UKSA) under grants ST/X002349/1 and UKRI888.

\section*{Data Availability}
The data underlying this article will be shared on reasonable request to the corresponding author.

\bibliographystyle{mnras.bst}
\bibliography{references}

\end{document}